\documentclass[letterpaper,english,aps,prl,twocolumn,floatfix,showpacs,amsfonts,amssymb]{revtex4}

 \usepackage{graphics}
\newcommand{\bq}{{\bf q}}
\newcommand{\bk}{{\bf k}}
\newcommand{\bp}{{\bf p}}
\newcommand{\bx}{{\bf x}}
\newcommand{\ba}{{\bf a}}
\newcommand{\be}{{\bf e}}

\begin{document}

\title{Coulomb interaction, ripples, and the minimal conductivity of graphene}

\author{Igor F. Herbut$^1$, Vladimir  Juri\v ci\' c$^1$, and Oskar Vafek$^2$}

\affiliation{$^1$ Department of Physics, Simon Fraser University,
 Burnaby, British Columbia, Canada V5A 1S6 \\
 $^2$ National High Magnetic Field Laboratory and Department of Physics, Florida State University, Tallahassee, Florida 32306, USA}

\begin{abstract} We argue that the unscreened Coulomb interaction in graphene provides a positive,
universal, and logarithmic correction to scaling of zero-temperature
conductivity with frequency. The combined effect of the disorder due
to wrinkling of the graphene sheet and the long range
electron-electron interactions is a finite positive contribution to
the dc conductivity. This contribution is disorder strength
dependent and thus {\it non-universal}. The low-energy behavior of
such a system is governed by the line of fixed points at which both
the interaction and disorder are finite, and the density of states
is exactly linear.  An estimate of the typical random vector
potential representing ripples in graphene brings the theoretical
value of the minimal conductivity into the vicinity of  $ 4 e^2 /h$.
\end{abstract}

 \maketitle

\vspace{10pt}

  Graphene, an atom-thick layer of graphite, has recently defined a new
frontier of condensed matter physics. Its essential electronic
property, inherent to the honeycomb lattice formed by the carbon
atoms, is that the low-energy quasiparticle excitations can be
thought of as being massless Dirac fermions, which propagate with a
Fermi velocity of around three thousandths of the velocity of light.
This pseudo-relativistic nature of the quasiparticle excitations
makes the electronic properties of graphene fundamentally new in
many respects  \cite{geim}. When the chemical potential is tuned to
the Dirac point graphene provides a rare example of a critical
two-dimensional fermionic system \cite{fradkin}. In contrast to its
textbook bosonic equivalent \cite{herbut3}, all sufficiently weak
interactions between electrons, including the long-ranged Coulomb,
are then irrelevant perturbations \cite{gonzales}, \cite{herbut1}.
The effects of electron interactions thus become progressively less
important as the system is probed at lower frequencies and
temperatures. One important consequence of this ``infrared freedom" is that the zero-temperature dc conductivity of
clean graphene is finite and universal, and simply determined by its
gaussian value of $\sigma= (\pi/2) e^2/h$ \cite{ludwig}. Including
scattering of impurities in a self-consistent Born approach yields
another, similar in magnitude and still universal,  value of
$(4/\pi) e^2/h$ \cite{fradkin}, \cite{ludwig}, \cite{shon}.
Localization corrections are also expected to set in at very low
temperatures \cite{aleiner} and thus further diminish the
conductivity. The actual measurements of graphene's conductivity,
however, are in significant discord with these results:
experimentally, $\sigma\approx 4e^2 /h$, and thus significantly {\it
larger} than all the theoretical values \cite{geim}. The origin of
this discrepancy is unclear at the moment, with several recent works
focusing on the role played by the {\it extrinsic} charged
impurities \cite{kumazaki}.

  In this Letter we show that the long-range Coulomb interaction between electrons in
  graphene provides the leading correction to the gaussian value of
conductivity, which is {\it positive} and by itself only logarithmically
slowly vanishing when frequency approaches zero.
This suggests that the origin of the observed unusually large
conductivity at the Dirac point may be {\it intrinsic}, and
originate from the Coulomb correction which is then effectively cut
off by a finite temperature/disorder/size effects.

 \begin{figure}[t]\label{flowFig}
{\centering\resizebox*{53mm}{!}{\includegraphics{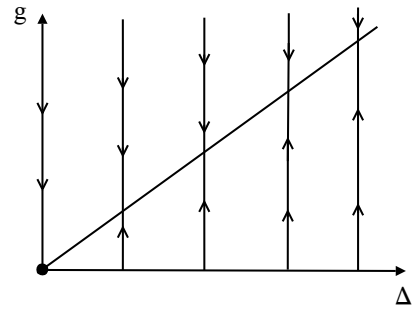}}
\par} \caption[] {The flow of the disorder strength $\Delta$ characterizing ripples of the
graphene sheet and the Coulomb coupling $g$ at weak
couplings. Two effects are balanced at the attractive line of fixed
points that emanates from the gaussian fixed point at the origin. At
the line the system exhibits linear density of states and a finite
non-universal dc conductivity.}
\end{figure}

We consider a
specific mechanism of such cutoff which invokes a non-trivial
interplay between long-range interactions and disorder
\cite{herbut2}, the latter being presently provided by the
apparently unavoidable wrinkling of the graphene sheet. The
combination of the Coulomb interaction and the random vector
potential that may be used to represent such ripples in graphene
leads to a line of stable fixed points (see Fig. 1). The finite
zero-temperature dc conductivity, obtained here from the Kubo
formula, varies continuously along the line and, most importantly,
{\it increases} with the increasing disorder strength.
  We provide the symmetry arguments for the existence of the line of stable fixed points,
and extract the density of states at low energies.  A crude estimate of the typical
parameters in graphene gives a sizable correction to the gaussian
value already to the lowest order in our calculation and
significantly narrows the gap that presently exists between the
theory and the experiment. Further implications of our theory
 and the connections with the critical bosonic
theories and related theoretical results in literature are
discussed.

The low-energy excitations in the vicinity of the Dirac points at
$\sigma \vec{K}$, $\sigma=\pm 1$ may be represented by two
four-component Dirac spinors $\Psi_\sigma ^\dagger  =
( \Psi_{\sigma \uparrow} ^\dagger (\bx,\tau), \Psi_{\sigma \downarrow}
^\dagger (\bx ,\tau))$. $\Psi_{\sigma \uparrow}$ is a two-component Grassman
field representing the components of the electron with the third
component of its (real) spin up on the two sublattices of the
honeycomb lattice and with wavevectors near $\sigma \vec{K}$
\cite{gusynin}.  The imaginary-time Lagrangian density for the
interacting system of quasiparticles in presence of a random vector
potential representing ripples of the graphene sheet
\cite{ludwig} is then
    $L=L_0 + L_C + L_D$, where
    \begin{equation}
L_0= {\bar\Psi}_\sigma^\alpha\gamma_{\mu}\partial_\mu\Psi_\sigma^\alpha,
\end{equation}
\begin{equation}\label{coulomb}
L_C=-ia_0^\alpha{\bar
\Psi}_\sigma ^\alpha\gamma_0\Psi_\sigma ^\alpha+a_0^\alpha\frac{|{\vec\nabla}|}{2g}a_0^\alpha,
\end{equation}
and
\begin{equation}\label{disorder}
L_D=-i\sigma{\bar\Psi}^\alpha_\sigma\gamma_n\Psi^\alpha_\sigma
A_n(\bx )+\frac{1}{2\Delta}A^2_n(\bx).
\end{equation}
Here the index $\alpha=1,2... , N$ labels replicas introduced to average over disorder, $\mu=0,1,2$, and $n=1,2$, and
$\{\gamma_\mu, \gamma_\nu \}=2 \delta_{\mu \nu}$.
The summation over repeated indices is assumed and the limit $N\rightarrow 0$ is to be taken at the end.
The integration over $a_0 (\bx,\tau)$ reproduces the standard $\sim g /|\bx |$ electron-electron interaction,
whereas integrating out the {\it static} vector potential  $A_n (\bx )$ yields an alternative form of $L_D$ \cite{ludwig}:
\begin{equation}
\tilde{L}_D=\frac{\Delta}{2}\int
d\tau'(\sigma{\bar\Psi}_\sigma^\alpha\gamma_n \Psi^\alpha_\sigma)(\bx ,\tau)(\sigma'{\bar\Psi}_{\sigma'}^\beta\gamma_n \Psi^\beta_{\sigma'})(\bx,\tau'),
\end{equation}
which will be also used. For convenience, we have set $\hbar=e/c=
v_F =1$, where $v_F\approx 10^6 m/s$ is the Fermi velocity. In this
convention there are two {\it dimensionless} coupling constants in
the theory: the Coulomb interaction $g=2\pi e^2/\epsilon\hbar v_F$, and the strength of disorder $\Delta$.

  In analogy with the two-dimensional bosonic critical systems \cite{herbut3}, at $T=0$ the 
conductivity at frequencies well below the microscopic energy scale  $\Omega/b $
in the units of $e^2 /h$ can be written in the scaling form as
\begin{equation}
\sigma(\omega) = F(b \omega, g (b), \Delta(b) ),
\end{equation}
where $F(x,y,z)$ is a {\it universal} scaling function. The functional dependence of
$g(b)$ and $\Delta(b)$ ensures the ultimate
independence of $ \sigma (\omega) $ on the arbitrary factor $b$.
We have set all the irrelevant couplings to zero.
Let us choose then $b\omega=\Lambda \ll \Omega $, with $\Lambda $ as an
arbitrary scale. Then
\begin{equation}
\sigma(\omega) = F(\Lambda, g ( \Lambda/\omega ), \Delta( \Lambda/\omega ) ).
\end{equation}
If the couplings $g ( \Lambda/\omega )$ and $\Delta( \Lambda/\omega )$
flow to small values as $\omega\rightarrow 0$, one can expand the
function $F(\Lambda, g,\Delta )$ as
\begin{equation}
\sigma(\omega)= F(\Lambda,0,0) + u g \left( \Lambda/\omega
\right) + w \Delta\left(\Lambda/\omega\right),
\end{equation}
to the first order in the two coupling constants.
$u$ and $w$ are constants. Kubo formula yields
the value of the gaussian term $F(\Lambda,0,0)= \pi/2$ in the
continuum limit \cite{ludwig} (see also Eq. (16) below). In what follows
we compute the remaining two terms in the last expression.

   \begin{figure}[t]
{\centering\resizebox*{70mm}{!}{\includegraphics{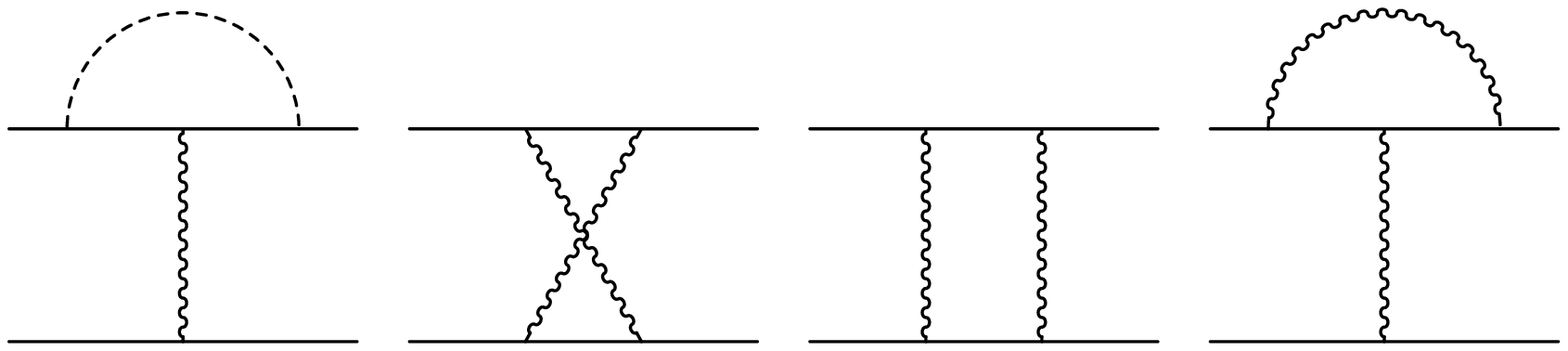}}
\par} \caption[] { The one-loop corrections to the disorder vertex in Eq. (4).
The dashed and wiggly lines stand for the scalar and the vector field
propagators, respectively. The last three diagrams sum up to zero. }
\end{figure}

  Let us first obtain the cutoff-dependent couplings $g(b)$ and $\Delta(b)$.
Non-analitycity in momentum of the second term in $L_C$ implies that
the flow  of $g$ with the change of cutoff can be written {\it exactly}
\cite{herbut2} as,
  \begin{equation}
  \frac{dg}{d\ln(b)} = (z-1) g,
  \end{equation}
  with the dynamical exponent $z$ fixed by the requirement $b^{z-1} =
Z_\omega/Z_k$. $Z_\omega$ and $Z_k$ are the wave-function and the
velocity renormalizations, respectively. To the first order in both
couplings one finds
  \begin{equation}
Z_\omega=1+\frac{\Delta}{\pi}\ln b,\qquad
Z_k=1+\frac{g}{8\pi}\ln b,
\end{equation}
in agrement with previous calculations \cite{gonzales},
\cite{herbut1}, \cite{ludwig}, \cite{stauber}, \cite{vafek}. At
$g=0$, the disorder strength $\Delta$ is exactly marginal coupling
\cite{ludwig}. We have also confirmed this by an explicit
calculation to order $\Delta^2$ (Fig. 2). When $g\neq 0$,
$\Delta(b)= \Delta Z_\Delta / Z_k ^2$. Using the form in Eq. (4) and
computing the first diagram in Fig. 2 then gives
\begin{equation}
Z_\Delta = 1+\frac{g}{4\pi} \ln(b).
\end{equation}
Thus, as the ultraviolet cutoff in the theory is changed from $\Omega$
to $\Omega/b$, the Coulomb and the disorder couplings flow
according to the differential equations \cite{stauber}
\begin{eqnarray}
  \frac{dg}{d\ln(b)} &=& g \left(\frac{\Delta}{\pi}- \frac{g}{8\pi}  +\mathcal{O}( g^2,\Delta^2, g\Delta) \right),\\
\frac{d\Delta}{d\ln(b)} &=& 0\label{deltaBeta}.
\end{eqnarray}
Under renormalization the electron interaction may thus both decrease
 or increase, depending on disorder.

Although the calculation has been performed here only to the leading
order, we suspect that the equation (\ref{deltaBeta}) may in fact be exact. Without the last
term in $L_D$ the rest of the (interacting) Lagrangian $L$ enjoys the
symmetry under the time-independent gauge transformation
\begin{equation}
A_n({\bx})\rightarrow A_n({\bx})+\partial_n\chi({\bx}), \qquad
\Psi_\sigma^\alpha \rightarrow e^{i\sigma\chi({\bx})}\Psi_\sigma ^\alpha.
\end{equation}
This implies that the usual Ward identities hold and that the polarization of the vector field $A_n (\bx)$ is transverse. The coupling $\Delta$, which in $L_D$ appears as the inverse mass for $A_n (\bx)$, should therefore not renormalize \cite{collins}.

 For completeness let us note that $L$  is also symmetric
under the purely time-dependent gauge transformation
\begin{eqnarray}
a_0 ^\alpha (\bx,\tau)&\rightarrow& a_0 ^\alpha (\bx ,\tau)+\partial_\tau f^\alpha (\tau), \\ \nonumber \Psi_\sigma
^\alpha &\rightarrow& e^{i f^\alpha (\tau)}\Psi_\sigma ^\alpha,
 \end{eqnarray}
 which guarantees the preservation of the form of $L_C$ and is ultimately
responsible for  Eq. (8) \cite{herbut3}.

 The conductivity to the first order in $g$ and $\Delta$ may be computed next.
We couple the external electromagnetic vector potential $\ba $
minimally to Dirac fermions and choose $\ba  = a {\hat \be} _1$,
for example. Using the Kubo formula for the replicated theory
\cite{herbut6} and to the first order in the two couplings we find
 \begin{equation}
 F(-i \omega, g, \Delta) = I_G (\omega) + g I_C(\omega)  + \Delta I_D (\omega),
 \end{equation}
 where the gaussian value is
 \begin{eqnarray}
 I_G (\omega) = 16 \pi \frac{d}{d\omega}  \int \frac{dq_0 d^2 \bq}{(2\pi)^3}
 \frac{ q_1 ^2 - q_2 ^2- q_0(q_0 - \omega)}{ q^2 ( (q_0 -\omega)^2 + \bq ^2)  } \\ \nonumber
  = \frac{\pi}{2} + \mathcal{O}\left(\frac{\omega}{\Omega}\right).
 \end{eqnarray}
 To the first order in disorder only the self-energy diagram contributes and yields
 a negative contribution
 \begin{eqnarray}
 I_D (\omega) = 64 \pi \frac{d}{d\omega} \int \frac{d\nu d^2\bk d^2\bp }{(2\pi)^5}
 \frac{ \nu (\nu^2-\bp^2) }{ (\nu ^2 +\bk^2)
 (\nu ^2 +\bp^2)^2 }\\ \nonumber
 \times \frac{(\nu+\omega )}{ (\nu +\omega) ^2 + \bp^2 } = -\frac{1}{6} + \mathcal{O}\left(\frac{\omega}{\Omega}\right).
 \end{eqnarray}
 Finally, the Coulomb contribution is
 \begin{eqnarray}
 I_C(\omega) = \frac{d}{d\omega}  \int \frac{d^2\bk d^2\bp}{(2\pi)^3 }
 \frac{4 \hat{\bk}\cdot \hat{\bp}}{ |\bk+\bp| (\omega^2 + 4 \bk^2) }
 \\ \nonumber
\times \left\{ \frac{\omega^2 - 4 \bk^2}{ \omega^2 + 4 \bk^2} -
\frac{\omega^2 +4 \bk\cdot \bp}{\omega^2+4\bp^2} \right\}
 = \left(\frac{25}{48}-\frac{\pi}{8}\right)+ \mathcal{O}\left(\frac{\omega}{\Omega}\right).
 \end{eqnarray}

  Few comments are in order at this point. The derivatives with
 respect to frequency in Eqs. (16)-(18), which are to be taken for
strictly positive frequencies, serve to subtract the finite
$\omega=0$ contributions to the integrals. These are known to arise as the artifacts of the
ultraviolet cutoff, which violates gauge invariance. Second,
in contrast to the disorder contribution, the Coulomb term contains
both the self-energy and the vertex corrections, given by the first
and the second term in the curly bracket in Eq. (18), respectively.
While each of these two separately is logarithmically divergent in
the continuum limit  $\Omega \rightarrow \infty$, the divergences
cancel out {\it exactly} in the full expression for the
conductivity. The final result in Eq. (18) represents the finite
remnant left after this cancelation. Indeed, the cancelation of
logarithms in the conductivity is to be expected: without it the
field-theoretic result would not be cutoff-independent, and would be physically meaningless.
The results in Eqs. (16)-(18) are universal numbers characteristic of
the continuum limit.

The last three expressions together with Eqs. (7), (11), and (12) at
finite disorder then give the dc conductivity
    \begin{equation}
    \sigma(0) = \left[\frac{\pi}{2} + (4 -\pi) \Delta + \mathcal{O}(\Delta ^2)\right] \frac{e^2}{h},
    \end{equation}
  and non-universal. The result is {\it larger} than the Gaussian value due to
the positive Coulomb contribution at the line of fixed points.  Note
that $\Delta$ in the last expression is the same as the ``bare" value of
the disorder strength at the microscopic scale.

In the ideal sample with $ \Delta =0$, on the other hand, solving the Eq. (11)
and inserting into Eq. (7) yields
\begin{equation}
\sigma(\omega) = \frac{\pi}{2} + \frac{\pi ((25/6)
-\pi)}{\ln(\Lambda /\omega) } +
  \mathcal{O}\left( \frac{1}{g \ln^2(\Lambda /\omega) } \right),
  \end{equation}
  in the limit $\omega\rightarrow 0$. The bare value of $g$
cancels out in the second term. Electron interactions provide therefore
the leading universal logarithmic correction to scaling of
conductivity at low frequencies in this case. Similar logarithmic corrections
arise in perturbative quantum chromodynamics, for example.

     To get an estimate of the size of the first-order correction  to
conductivity in Eq. (19) we restore all the constants  we previously
set to unity. For the dielectric constant of $\epsilon
\approx 6$ we find $g\approx 2$ \cite{schedin}. The conductivity is determined by
the bare value of disorder, not the interaction, however. The
disorder coupling can be written as
     $\Delta^{1/2}\approx \Phi_\xi/\Phi_0$,
     where $ \Phi_\xi\approx h \xi^2$ is the flux of the effective random magnetic field $h= \partial_1 A_2 - \partial_2 A_1$ through the ripple of the average size $\xi$, and $\Phi_0= \hbar c/e$ is the flux quantum.
     Effective magnetic field $h$ can be estimated by assuming that the same random
magnetic field is responsible for the observed suppression of weak
localization in graphene \cite{morozov}. Using $h\approx 1 T$ and
$\xi \approx 30 nm$ \cite{morozov} we find $\Delta
\approx 2$, and $\sigma\approx 3 e^2/h$, to the leading order.
Clearly this is only a crude estimate and for $\Delta$ of order
one the higher-order terms in Eq. (19) need to be included. It is
encouraging, however, that the size of the lowest-order correction
is significant and in the direction towards the experimental result.
The latter result is a direct and a non-trivial consequence of the
unscreened long-range Coulomb interaction in graphene. It may also be relevant
 that the experimental observation of
normal localization properties in graphene seems to correlate with a lower
minimal conductivity \cite{novoselov}. The present theory would
naturally account for this since both effects result from a low
value of $\Delta$.

   A limitation of our result should also be pointed out. We have computed the
$T=0$, $\omega\rightarrow 0$  conductivity along the line of fixed
points, whereas the measurements of the minimal dc conductivity in
graphene typically correspond to the opposite, $\omega=0$,
$T\rightarrow 0$ limit. While both conductivities at a fixed
disorder $\Delta$ are expected to be universal, the two
numbers could in principle be different \cite{damle}. If the results
from the related critical bosonic theories with \cite{herbut5} and
without disorder \cite{damle}, \cite{herbut3} are of any guide, the latter universal
number may be expected to be only larger. Its computation along the
line of the fixed points at which both the disorder and interactions
are finite may be a non-trivial task though.

   Our result at $\Delta=0$ is in stark contrast to the recent result in
ref. \cite{mishchenko}, where Coulomb interaction is claimed to
suppress the conductivity at low frequencies. We note that this
conclusion is in contradiction to the well-established infrared
irrelevancy of the Coulomb coupling \cite{gonzales}, \cite{herbut1},
which implies that the Coulomb interaction can only provide
corrections to the Gaussian conductivity, which ultimately vanish in
the dc limit. Furthermore, the result of ref. \cite{mishchenko}
is explicitly dependent on the cutoff in the Dirac
theory, which is completely arbitrary. In contrast, the cancelation
of the two logarithmically divergent terms \cite{schmalian} in Eq. (18) guarantees
the renormalizability of our result. The low-frequency conductivity we
computed is consequently perfectly cutoff-independent to the order
of our calculation, as it has to be if the picture of Dirac
quasiparticles is to remain physically meaningful in presence of the
interactions.

Finally, let us adress the density of single-particle states: 
$N(\omega )\propto \omega ^{ (2-z)/z }$, just as at the unstable line at $g=0$. Since $g\neq 0$
however, Eq. (8) implies $z=1$ at the stable line. The density of
states is thus exactly linear, in contrast to the $g=0$ line.

To summarize, our main finding is that the lowest-order combined
effect of electron interactions and rippling in graphene is to
increase its minimal dc conductivity in a non-universal,
disorder-dependent fashion. A testable prediction of our theory
would be a decrease of minimal conductivity in graphene with the
suppression of wrinkling, which, incidentally, should also
 bring back the usual localization behavior at finite density.

I.F.H. and V.J. are supported by the NSERC  of Canada. I.F.H. and O.V.
are grateful to KITP at UC Santa Barbara (NSF grant
PHY99-07949) for its hospitality during its graphene workshop at
which this work was initiated. O.V. wishes to acknowledge useful
discussions with Dr. M.J. Case.

\end{document}